\documentclass[a4paper]{article}

\usepackage[margin=1in]{geometry} 

\usepackage{amsmath}
\usepackage{float}
\usepackage{ragged2e}
\usepackage[verbose]{wrapfig}
\usepackage{amsthm}
\usepackage{appendix}
\usepackage{fancybox}
\usepackage{colortbl}
\usepackage{amssymb}
\usepackage{array}
\usepackage{pgfplots}
\usepackage{tkz-berge}
\usepackage{tabularx}
\usetikzlibrary{arrows,shapes}
\usepackage{dot2texi}
\usepackage[printwatermark]{xwatermark}
\usepackage{xcolor}
\usetikzlibrary{shapes,arrows}
\usetikzlibrary{shapes.multipart, calc}
\usepackage{lipsum}
\usepackage[utf8]{inputenc}
\usetikzlibrary{arrows,automata}
\usetikzlibrary{positioning}
\tikzset{
    state/.style={
           rectangle,
           rounded corners,
           draw=black, very thick,
           inner sep=2pt,
           text centered,
           },
}

\tikzstyle{block} = [draw, rectangle, 
    minimum height=3em, minimum width=6em]
\tikzstyle{sum} = [draw, circle]
\tikzstyle{input} = [coordinate]
\tikzstyle{output} = [coordinate]
\tikzstyle{pinstyle} = [pin edge={to-,thin,black}]

\usepackage[utf8]{inputenc}
\usepackage{hyperref}
\hypersetup{
	unicode,
	pdfauthor={Chase Smith, Alex Rusnak},
	pdftitle={Networking},
	pdfsubject={},
	pdfkeywords={},
	pdfproducer={LaTeX},
	pdfcreator={pdflatex}
}

\usepackage[sort&compress,numbers,square]{natbib}
\pgfmathdeclarefunction{gauss}{2}{%
  \pgfmathparse{1/(#2*sqrt(2*pi))*exp(-((x-#1)^2)/(2*#2^2))}%
}

\theoremstyle{plain}
\newtheorem{theorem}{Theorem}

\theoremstyle{definition}
\newtheorem{definition}[theorem]{Definition}

\usepackage{graphicx, color}
\graphicspath{{./fig/}}

\usepackage{graphicx}

\usepackage{algorithm, algpseudocode} 
\usepackage{mathrsfs} 

\usepackage{lipsum}

\title{Proximity-based Networking: Small world overlays optimized with particle swarm optimization}

\author{Chase Smith \\ \texttt{chase@proxima.one, chasesmith@berkeley.edu}
   \and Alex Rusnak \\ \texttt{alex@proxima.one} }

\begin{document}
\pgfdeclarelayer{background}
\pgfdeclarelayer{foreground}
\pgfsetlayers{background,main,foreground}

\begin{figure}[H]
\begin{center}

	\maketitle

	\begin{abstract}
	
Information dissemination is a fundamental and frequently occurring problem in large, dynamic, distributed systems. In order to solve this, there has been an increased interest in creating efficient overlay networks that can maintain decentralized peer-to-peer networks. Within these overlay networks nodes take the patterns of small world networks, whose connections are based on proximity. These small-world systems can be incredibly useful in the dissemination and lookup of information within an internet network. The data can be efficiently transferred and routing with minimal information loss through forward error correct (FEC) and the User Datagram Protocol (UDP). Cite.

	We propose a networking scheme that incorporates geographic location in chord for the organization of peers within each node's partitioned key space. When we combine this with a proximity-based neighborhood set {based on the small world structure} we can mimic the efficient of solutions designed to solve traditional small-world problems, with the additional benefit of resilience and fault-tolerance. 
	
	Furthermore, the routing and address book can be updated based on the neighborhood requirements. The flexibility of our proposed schemes enables a variety of swarm models, and agents. This enables our network to as an underlying networking model that can be applied to file-sharing, streaming, and synchronization of networks.  
	\end{abstract}
\end{center}
	\end{figure}
	\tableofcontents
 
 \section{Introduction}
 
 Peer to peer networks are increasing in popularity as an efficient way to propagate information, search file systems and stream data. Prevalence of low-cost data transmission methods like UDP and QUIC enable data streaming from one client to many \cite{iyengar2018quic, postel1980rfc0768}. When combined with WebRTC, it is possible to enable browser clients to act as peers within the network \cite{nurminen2013p2p}. The implementation of peer-to-peer networks have several advantages over the traditional server-client relationship. They enable a server to have lower communication cost, and much greater scalability. Further, they provide greater fault-tolerance, and autonomy within a network. In exchange for these advantages, peer-to-peer networks give up guarantees on latency, overall network communication costs, and add complexity to a network. Despite these drawbacks, peer-to-peer networks are presently utilized for large data transfer, file storage, and within blockchains.
 Current implementations of peer-to-peer networks lack spatial awareness of the underlying peers, this causes the network to be partitioned based on random hashes of the data instead of the locality of the peer. Implementations of overlay networks like BitTorrent, IPFS, and even Bitcoin utilize this method of networking and organization \cite{cohen2003incentives}, \cite{benet2014ipfs}, \cite{nakamoto2019bitcoin}.

Our proposed solution is to provide an underlying peer network that utilizes the Chord distributed hash table, with a key hash that is based on the location of the peer within the node. This method seen in \cite{pethalakshmi2014geo}, provides efficient routing within a network, as well as a deterministic routing table for each node. Additionally, our network incorporates a dynamic, agent-based selection scheme for each node's neighborhood selection that can be configured in an independent manner. The complete network provides efficient routing, peer autonomy, and flexibility as an underlying network for file-sharing, blockchain, and data streaming overlays. 
\newpage
\section{Preliminaries}

\subsection{Networks}
Networks are composed of vertices representing nodes, and edges representing connections. This relationship can be represented as both a matrix or a graph $\mathbb{G}(V, E)$ with $V$ denoting peers, and $E$ denoting connections. There are a variety of applications for networks and graphs, one major application is networking within computer networks over the internet. Graphs are represented as a set of vertices that are connected by edges. There are a variety of use cases. Computer networks over the internet are used for several use cases:

\begin{itemize}
    \item \textbf{Data lookup}. Data lookup and addressing can be done in a variety of ways, generally relying on distance metrics between ID string.
    \item \textbf{Peer lookup}. Peer lookup can be done through an overlay network, or through flooding within a system in a similar manner to Data lookup.
    \item \textbf{Information Synchronization}. Distributed ledgers, like Blockchains, maintain networks designed to store and update the same set of information. In these cases networks are motivated to attain the lowest synchronization time. 
\end{itemize}

\subsection{Data Transmission} 
There are a variety of methods for communicating and passing data between nodes in the network. Each of these has its own benefits, but User Datagram Protocol (UDP) is often the favored approach because of its flexibility.
\begin{itemize}
    \item \textbf{Transmission Control Protocol (TCP)}. The Transmission Control Protocol relies on the communication and verification from both communicating parties.  
    \item \textbf{User Datagram Protocol (UDP)}. The User Datagram Protocol (UDP) is an alternative to the Transmission Control Protocol. It does not require the confirmation from the receiving party, and thus can be done in a faster and more efficient manner. Furthermore, implementation of UDP like Quick UDP Internet Connections (QUIC) have fixed issues with security and provides a two-party handshake system that mimics TCP.
\end{itemize}

\subsubsection{Forward Error Correction (FEC)}
The User Datagram Protocol (UDP) suffers from the loss of information during data transfer. In order to address this issue, information is encoded such that a subset of a message can be used to recreate the entire message. The method of encoding these messages is known as forward error correction (FEC) \cite{biersack1992performance}. Through FEC, encodes each message $m$ with an extra $k$ bits, such that if a message losses $k$-1 bits, it can still be recreated.

\subsection{Overlay Networks}
Overlay networks require a method for routing, hashing, and neighbor selection. Below are some examples.  Kademlia, Chord, and Gnutella are examples of such overlay networks. They require a logical organizational method that accounts for Peer selection, Routing of messages, and distributed hash table. In return, they provide certain attributes of: 
\begin{itemize}
\item \textbf{Autonomy and decentralization}. The nodes collectively form the system without any central coordination.
\item \textbf{Fault tolerance}. The system should be reliable (in some sense) even with nodes continuously joining, leaving, and failing.
\item \textbf{Scalability}. The system should function efficiently even with thousands or millions of nodes. 
\end{itemize}
Overlay networks share the advantages of resilience and being able to broadcast messages, but they suffer from duplicate messages, long latency, and the slow spread of data within the network. 

\begin{definition}{\textbf{Distributed hash table.}} is a distributed system that provides a lookup service similar to a hash table[1]: (key, value) pairs are stored in a DHT, and any participating node can efficiently retrieve the value associated with a given key. The main advantage of a DHT is that nodes can be added/removed with minimum work around re-distributing keys. Keys are unique identifiers which map to particular values, which in turn can be anything from addresses, to documents, to arbitrary data.
\end{definition}

\subsubsection{Key space}
After the method for key creation has been established, there must be an established rule for dividing the keyspace of the network. The partitioning is based on some distance metric (XOR, Ring, etc) with each scheme providing its own methods for routing and peer selection. 
\begin{quote}
\begin{itemize}
    \item \textbf{Consistent Hashing}.  Hashing that provides a consistent response to whatever the peer id is, regardless of the location, spatial awareness, etc. 
    \item \textbf{Locality-preserving}. Hash function that takes into account the location of the nodes, and can be optimized for proximity-based peer selection. 
\end{itemize}
\end{quote}

\begin{itemize}
    \item \textbf{Routing}. Routing is done through the DHT itself, it is meant to optimize load, bandwidth, and latency within the network.
    \item \textbf{Peer Selection}. The selection of peers is based on the organization of the DHT itself. Generally, each node carries a table of peers that has some relation to the node itself. Whether it is proximity, distance, or ring-based. 
\end{itemize}

\subsection{Dual Convex Optimization}
 Overlays can be designed for a variety of use-cases, similarly they can be solved using a variety of methods including linear programming, reinforcement learning, evolutionary algorithms, and particle swarm optimisations. The value of this method is that it can be dynamic, is constantly evolving, and allows for agents to be independent with low computational overhead. Sub-problems within overlays can be optimized through a variety of methods as well. 
\begin{itemize}
    \item \textbf{Objective Function}. The objective function is designed to be optimized via minimization or maximisation. Additionally, it is possible to include multiple objective functions in the form of vector. $\langle$ $f_1(x)$, ..., $f_m(x)$ $\rangle$ to be evaluated. where $f_i(x)$ represents the i$^{th}$ objective function.
    \item \textbf{Constraints}. Additional constraints can be added into the optimization function to account for constraints to accommodate for requirements of the optimal solution. The requirements represent the necessary rules that state must adhere to, and vary based on each node. 
\end{itemize}

\subsection{B$^*$ Search} 
This search uses a best-first-approach that combines Breadth First Search with a weighted path problem, and a heuristic estimation. $B^*$ searches excel at finding nearly optimal paths within uncertain networks. As an optimization methodology, it can applied to routing decisions with an overlay network.
\begin{quote}
    $$f(n) = g(n) + h(n)$$
    \\$g(n)$ represents the cost function of a hop from the source to $n$ 
    \\ $h(n)$ represents the heuristic estimation of the cost of $n$ to the source
\end{quote}

\subsection{Particle Swarm Optimization} 
Particle Swarm Optimization belongs to a set of optimization that algorithms that use velocity, positioning, and additional parameters to solve a local and global optimum problem \cite{kennedy1995particle}.
  Our network engages a Particle Swarm Optimization methodology in which each node acts as its own particle, with its own set of candidate solutions for each level in the hierarchy. A node stores two solutions: a global solution that represents the best position that a node has found, and a candidate solution which is the current position that a node is exploring. 
 \begin{itemize}
     \item \textbf{Position}. This represents the specific solution that a particle holds, it is described as a vector of values: $\langle$  $v_1$, ... , $v_n$ $\rangle$. 
     \item \textbf{Velocity}. When each node updates its position it does so according to a velocity vector. 
 \end{itemize}
Nodes within this scheme cyclically change state, going from randomly sampling the network to validating their changes to communicating them to other nodes. As these nodes converge to a local optima they begin to slow down their update cycles, and thus the network overhead. 
  \begin{itemize}
     \item \textbf{Exploration.} Nodes that are in the \textit{Exploration} state conduct random samplings of the network, with the intention of evaluating nodes, and sharing information. The randomness of these exploration decisions allows for better updates and can break peers out of local optima. 
     \item \textbf{Diffusion.} Along with exploration, nodes can receive benefits from neighbors that have finished explorations. Once the network change is decided, the new state of the network can be broadcast to the impacted nodes. Other nodes can also sample updates from their neighbors.
    \item \textbf{Validation.}  Models are validated using random samplings of the current network. This can occur in different ways, with the fundamental idea being validation through law of large numbers. 
 \end{itemize}

 \subsection{Gaussian Mixture Models}
 Gaussian Mixture Models (GMM) combine multiple Gaussian distributions to represent probability distributions within an entire network \cite{reynolds2009gaussian}. Clustering can be done through the use of a dynamic particle swarm based algorithm. While in actuality clusters do not utilize a single radii to determine membership, but rather a probability density function , as seen in Gaussian Mixture Models. This means that clusters can be represented by a mean and standard deviation $\mathcal{N}(\mu ,\sigma ^{2})$, and it means that nodes can represent a combination of multiple clusters. 
     $$\sum_{i=0}^K \phi_i {\mathcal{N}_i}(\mu_i, \Sigma_i^2)$$

\subsubsection{Gaussian Distributions}
Gaussian distributions are used in optimization and classification problems because they effectively model the distribution of real-world events. Gaussian distributions are normal distributions [$\mu$, $\sigma$] that can be represented by a mean $\mu$, and a covariance matrix $\sigma$, both of  whose dimensionality is based on that of the data being modelled. This is represented with the equation:
$$\phi (x) = \frac {1}{\sigma \sqrt{2\pi}} e^{- (\frac{x-\mu }{2 \sigma })^{2}}$$
These distributions are estimated using the methodology seen in
After each estimation, they can be updated by moving towards the expectation maximization function. 

 \subsubsection{Maximum Likelihood Estimation}
 Gaussian Mixture Models are organized based on their heirarchical levels. They can be split and merged according to population, cost, or distance. This can be done by pre-selecting a number of parameters or requirements to govern the creation of clusters. Once the initial base level of clusters is determined these cluster need to be combined into $m$ clusters of a higher level. These Gaussian Mixtures are represented as a weighted combination of the underlying Gaussian Distributions. Gaussian distributions can be calculated and updated according to the maximum likelihood estimation-- towards the best match of the observed variables. This estimation is done for each level of clusters by optimizing the function for that specific cluster level.

\section{Related work}
There exist a variety of overlay networks and distributed hash tables that relate to our solution and are a part of the research knowledge. Kademlia exists as a distributed hash table scheme in which keys are partitioned based on the XOR metric \cite{maymounkov2002kademlia}.  IPFS, Bittorrent, and blockchain implementations like Ethereum have implemented this scheme \cite{benet2014ipfs, johnsen2005peer, wood2014ethereum}. 

Chord provides a ring-based distributed hash table, with distance metrics that are based on circular distance \cite{stoica2001chord}. Pastry is similar to chord in that it has a ring shaped key space. In addition to its routing list, each node contains Node IDs are chosen randomly and uniformly so peers who are adjacent in node ID are geographically diverse \cite{rowstron2001pastry}.

Chord has been implemented with a variety of hashing schemes including Geo-chord which bases the ring on geographic locations ($latitude$, $longitude$) \cite{pethalakshmi2014geo}. Additionally, there exists another chord scheme that uses proximity is PChord which bases node neighborhoods on the RTT between nodes. 
\cite{hong2006pchord}. Unfortunate;y, PChord does not discuss the key generation and partitioning method for the network.

\section{Our Contributions}
We use a combination of static Distributed Hash Tables (DHTs) along with a locality preserving key partitioning scheme.  Additionally, nodes construct a neighborhood of peers through proximity-based conditions. A node's neighborhood can satisfy various requirements and objective functions, and it can be optimized through multi-modal particle swarm optimization \cite{kennedy1995particle}.

This proposed solution uses a hybrid approach with static, locality-based Distributed Hash Tables (DHTs), as well as dynamic neighborhoods. This increases the storage and memory requirements for each node, but it provides more efficient routing , and greater amount of scalability for the network itself. 

Our proposed solution that enables efficient routing, dynamic autonomous, small hash tables, and anonymity, and fault tolerance. Because of its versatility, our solution can act as an underlying network that can be applied to file-sharing, message routing, information dissemination, and pub/sub architecture.

\section{Motivation}
To scale peer-to-peer applications it is necessary to optimize the efficiency, speed, and versatility of the underlying peer-to-peer network. In order to provide a maximal amount of performance it is important to provide an optimization framework whose parameters can be updated quickly, and that can evolve to deal with a dynamic network. Thus, the motivation for our paper is to provide an overlay network that features a level baseline performance that can be further tuned to match the constraints of a variety of networks, in a rapid manner. The eventual goal for this network to efficiently emulate the routing path that ISPs would use to route the packet from node to node. 

\section{Overlay network} 
We propose a hybrid peer-to-peer overlay network that is constructed with a key hashing scheme that is based on geographic location and a Chord Distributed Hash Table. Our Chord DHT partitions keys based on their location within in a hierarchical set of clusters instead of their exact locations. These clusters are recursively generated using Particle Swarm Optimization (PCO) and K-means clustering. The motivation for this is to have nodes placed in clusters based on geographic location with the ability to dynamically optimize their routing tables to accommodate for differences in Round Trip Times (RTT).
 
 Our proposed solution is designed with versatility in mind. The flexible nature of PSO enables for for the inclusion of additional network requirements to adapt to needs of fault-tolerance, resolution, and overall performance. This means that our network can adapt to applications that require low storage overhead, communication overhead, or path latency. 
\begin{quote}
\textbf{Ring-based Distributed Hash Table}. Node keys are hashed using a locality preserving hashing scheme that projects latitude and longitude into a single dimension. Keys are partitioned based on their proximity to clusters in the network. Clusters can be split or merged  through dual-optimization (Necessary to check) of PSO to optimize our recursive clustering scheme. \\ \\
\textbf{Routing}. Routing of packets is done through gradient descent, in which the distance and number of hops is meant to be minimized.  This can be done through a combination of greedy search, and path optimization through pheromone secretion along efficient paths. \\ \\
\textbf{Network Optimization}. Our network is optimized and updated using Particle Swarm Optimization that using two objective functions, one that is composed of requirements and one that is composed of an optimization function. The combination of these functions enables requirements to be added to the problem space, and imbues our network with an added level of versatility.
\end{quote}

\section{Ring-based Distributed Hash Table}

Our network topology is partitioned into a distributed set of hash tables using a recursive set of ring-based Chord distributed hash tables. Keys are hashed according to the location of the nodes within the grid, and then projected into a single dimension using a z-cantor set. This projection a locality-preserving implementation that adheres to the properties of ring. This enables keys to be partitioned based on location, and preserve locality-based distance as a metric for DHT organization. Despite being able to partition keys using location, nodes within the network are mainly organized by their proximity to clusters in the network.

\subsection{Locality Preserving Key Hashing} 
As stated previously, our network organizes node keys by geographic location. This requires several transformations of node location, as well as the potential need for a location approximation method. The location of the nodes is transformed from latitude and longitude to the USNG grid system (e.g. 18S UJ 23371 06519). This hash is then projected into a singe dimension so that its key can effectively be partitioned by location.

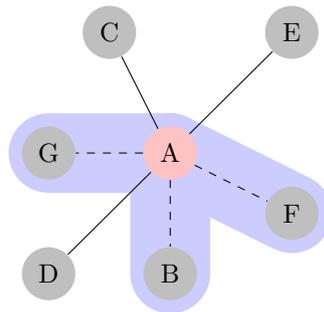
\begin{wrapfigure}{r}{0.35\textwidth}
\tikzstyle{vertex}=[circle,fill=black!25,minimum size=20pt,inner sep=0pt]
\tikzstyle{selected vertex} = [vertex, fill=red!24]
\centering
\begin{tikzpicture}[node distance=3cm, scale=0.8]
    
    \node [selected vertex, name=A] (A) at (2, 2) {A};
        \node [vertex] (B)  at (2, 0) {B};
              \node [vertex] (C)  at (1, 4) {C};
                     \node [vertex] (D)  at (0, 0) {D};
              \node [vertex] (E)  at (4, 4) {E};
                      \node [vertex] (F)  at (4, 1) {F};
              \node [vertex] (G)  at (0, 2) {G};
    \draw [-, dashed] (A) -- node {} (B);
    \draw [-] (A) -- node {} (C);
      \draw [-] (A) -- node {} (D);
        \draw [-] (A) -- node {} (E);
          \draw [-, dashed] (A) -- node {} (F);
            \draw [-, dashed] (A) -- node {} (G);
    
    \begin{pgfonlayer}{background}
        \draw[rounded corners=2em,line width=3em,blue!20,cap=round]
                (A.center) -- (B.center);
                                                 \draw[rounded corners=2em,line width=3em,blue!20,cap=round]   (A.center) -- (G.center);
                                                                         \draw[rounded corners=2em,line width=3em,blue!20,cap=round]   (A.center) -- (F.center);
                
\end{pgfonlayer}
\end{tikzpicture}
\caption{Location approximation through centroids or anchor nodes whose locations are known.}
\end{wrapfigure}

\subsubsection{Location Approximation}
While many nodes will be able to broadcast their own location honestly, there will be cases of malicious nodes sending inaccurate locations, or nodes unable to send their location. In these instances location can be approximated using other anchor nodes whose identities are known within the network. This approximation is done the centroid nodes within differently geographically partitioned regions, and using the Round Trip Time from each centroid to estimate its location. If nodes broadcast messages to the node joining the network its location can be approximated using the message round trip time (RTT) and the locations of these nodes. It is important to note that these centroid nodes can cheat, but this can be undone by polling the RTT for multiple anchor nodes, and then calculating based on these figures. If there exists any issue with this method it then becomes possible to approximate location using typical latitude, longitude, with malicious nodes being penalized.  Using a satellite approximation method, the location of a node can be ascertain. 

\begin{wrapfigure}{l}{0.35\textwidth}
\centering
\includegraphics[width=0.34\textwidth]{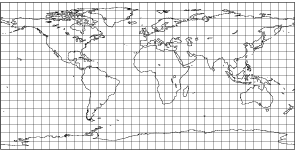}
\caption{The USNG Grid Organization}
\end{wrapfigure}

\subsubsection{Location Transformation}

 Once the location of the node is found, the node is grouped into one of the grids within the US grid system \cite{federal}. The advantages of this system over the typical latitude and longitude is that unlike latitude and longitude each unit in the grid has an equal distance, the units are already encoded in alpha numeric characters, and the location can be identified with a varying level of precision. The Grid system is then projected from two dimensions to a one dimension using the same technique seen in \textbf{CITATION}. This projection maintains the integrity of its distance metric and enables nodes to be organized and partitioned by geographic location. 

\subsubsection{Key Partitioning} 
Following the generation of the locality preserving key hash, the nodes are organized within the distributed hash tables. As stated previously, the nodes are partitioned based on the geographic location of the centroid within their clusters. Clusters are then identified within the hierarchy by their shortest unique prefix string.

\subsection{Peer Routing Table}
The routing table comprises of peers in different regions, with a key partitioning scheme that  constructs finger tables using a ring-based topology. The routing table for each node is designed to create a deterministic network organization that can be used in generic path selection. It is organized such that each node has a set of peers within a logarithmic number of steps. It does this by uses a ring-based topography with each level of the hierarchy acting as a new ring. This creates a logarithmic size peer routing table for each level. 
    $$\sum_{i=0}^h log_2(k^i)$$
A node's hash table is largely dependent on the topology of the network, and the distribution of clusters within the network. Nodes maintain peer tables that contain $h$ buckets where $h$ denotes the height of the heirarchical cluster. The buckets are filled by $log(k)$ nodes within each cluster level. This presents a similar structure to Kademlia, but instead of using the discontiguous XOR metric it uses rings at each level. This creates a contiguous state space that more accurately represents geographic regions.

\begin{wrapfigure}{r}{0.3\textwidth}
\centering
$\begin{Bmatrix}
c_{11} & ... & ... & ... & c_{1n}\\
... & ... & ... & ... & ...\\
... & ... & ... & ... & ...\\
... & ... & ... & ... & ...\\
c_{n1} & ... & ... & ... & c_{nn}\\
\end{Bmatrix}$
\caption{Adjacency matrix, where $c_{ij}$ denotes the optimal path between cluster centroid $i$ to centroid $j$}
\end{wrapfigure}

\subsubsection{Adjacency Matrices}
Along with peer routing information, each node stores an adjacency matrix that tracks the path latencies between clusters. Adjacency matrices are generated for each level within the routing table and provide information that can be used for individual node optimization, as well as global network optimization. These matrices are used when making routing decisions, and can be further supplemented with additional heuristic indicators and features. 

\subsubsection{Heuristic Indicators} 
Heuristic indicators can be attached to nodes within a Peer routing table to improve the effectiveness of route selection, and in routing table optimization. These indicators can include availability, load, latency, as well as usage statistics on messages being sent. This accumulated knowledge enables the node to determine areas in its routing table that can be updated to optimize the individual node's message distribution.

\section{Routing}
Routing within our network is designed to accommodate for the differences between geographic topology and the distribution of our global internet infrastructure.  
Since, data transmission within the internet infrastructure is primarily based on cable networks and physical infrastructure, messages that can move at different speeds in different locations. This means that geographic distance is not an ideal metric when determining optimal path selection because different geographies have differing latencies. Thus, while we partition and cache keys utilizing geography, we store clusters, and select paths, using estimated path latency. 
\\\\
Since routing is done between a variety of grid squares, from a Cantor Set projection, distance and RTT are the easiest weights to use, with the motivation being that internet connections are geographically placed to ensure an efficient flow of traffic between areas of high density.
\\\\
Our solution routes data through a greedy path selection scheme, where the next hop is chosen based on the shortest anticipated path length. Routing path selection is done using a greedy Best First Approach with a weighted path estimation algorithm.

\subsection{Routing Estimation} 
The route estimation is designed to be as accurate as possible, with the path latency of routes trending towards the optimal latency, as the node incorporates more information about the network and its own usage. \\
\begin{wrapfigure}{r}{0.45\textwidth}
\includegraphics[width=0.44\textwidth]{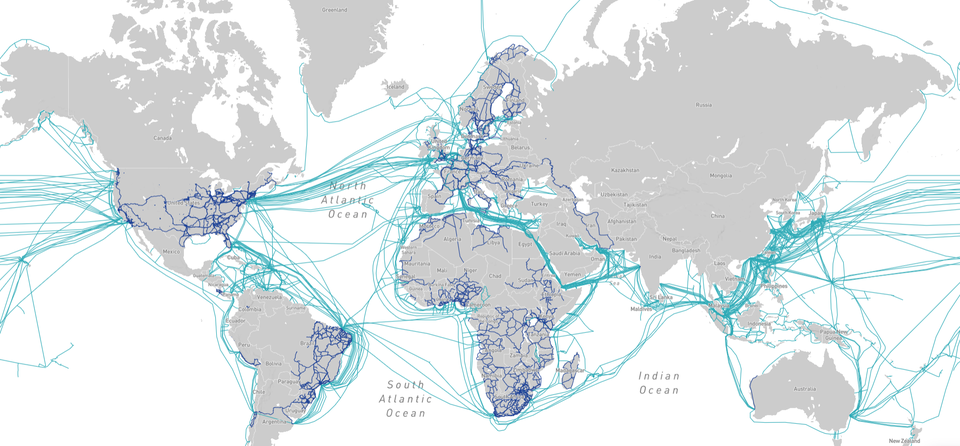}
\caption{Internet Physical Infrastructure. \scriptsize{(https://preview.redd.it/jy6459fajpy21.png)}}
\end{wrapfigure}
Rosutes are estimated using the node's adjacency matrix, as well as its understanding of the organization of cluster hierarchy within the network.  The adjacency matrix is important because it shows the optimal path between any two nodes, and allows the node to estimate the structure of the underlying internet infrastructure. \\\\ 
As stated previously, Geographic distance is not an ideal metric when determining optimal path selection because different geographies have differing latencies. 
\\\\
Instead, each node is given a set of clusters as well as their weighted path lengths in latency. This enables the node to estimate the velocity gradient of the entire surface area, as well as the geographic placement of nodes. With path selection being based on the estimated number of hops and the anticipated latency of each hop. 

\subsection{Best-First Search}
Routing paths are selected using a greedy best-first approach, with the next hop being selected based on the lowest estimated path latency. This estimation is calculated in part by looking at the latency gradient and the anticipated routing tables of the next nodes, and also by maximizing the number of "levels" in the cluster hierarchy that are traversed.
\begin{center}
\noindent\fbox{
\parbox{0.8\textwidth}{
   
       $$f(n) = g(n) + h(n)$$ \\
    
    \RaggedRight
    $g(n)$ represents the cost function of a hop from the source to $n$ \\
    $h(n)$ represents the heuristic estimation of the cost of $n$ to the source }
    }
\end{center}
Our networks route paths using a best-first search that uses a heuristic estimation function as well as a cost function to determine the shortest path. Paths are explored within each node in a Breadth First Search manner, with shortest paths being explored first, and the next hop being determined by the convergence of the shortest path. If there are issues with optimal convergence, additional objectives can be added such that path length is considered within the cost function. 
\subsection{Route Optimization} 
In order to reduce the number of paths that are explored, and thus reduce the complexity of the optimization problem, it is possible to compute and store estimated path weights. This reference table can then be optimized as the node develops a greater understanding of the network topology. 

While optimization estimates of the networks can be done through individual message transmissions, but they be built using state samplings that occur as a result of the continuous network optimization that nodes do. 

Optimization of the routing table can be folded into the general network optimization evaluation method, with samplings of network information being done based on usage, understanding of the network, and random selection. These samplings can then be used to update the estimated cost of routes. 

Additional routing optimizations will be done through the use of the \textit{traceroute} that can be done through traditional ISPs. This tracks the route of the message, and can be used to find the optimal paths between points. 

\section{Autonomous Network Organization}

Our network is optimized according to the accumulated work of individual nodes, in an autonomous organized optimization model, as seen in \cite{verbeek2003efficient, eckart2018hgmr}. Nodes within this scheme cyclically change state, going from randomly sampling the network to validating their changes to communicating them to other nodes. As these nodes converge to a local optima they begin to slow down their update cycles. 

As the network progresses, each node develops an understanding of the probability distribution of the nodes within the network, and can begin to set up Gaussian mixtures to fit a probability density function within the network space. These mixtures represent an ever greater number of nodes, with a wider latency reach. Exploration occurs in intervals, with nodes sampling peer routing tables from random peers within selected cluster levels.  These models agglomeratively generate a heirarchical set of Gaussian distributions to estimate network traffic or population density within different layers of mixture models.
These clusters are represented by centroids and interval defined areas, and are composed of all of the nodes that are encapsulated by this area. Nodes within the network start with a set of nodes in their routing table, that are  selected according to a random distribution. Since these Gaussians are organized in a hierarchical model, their will be a different solution for each level. They maintain a solution for each hierarchy. Exploration occurs in intervals, with nodes sampling peer routing tables from random peers within selected cluster levels.

\subsubsection{Network Estimation Models}
Nodes model both the population and traffic of the network using particle swarm optimized Gaussian Mixture Models. It should be noted that the networks can also be estimated without PSO, but it requires a larger communication overhead, and added time to convergence. Each node begins with a network estimation that mimics a uniform distribution, and builds up larger mixtures of Gaussian distributions. This methodology is used in the generation of two separate network estimations: traffic flow, and population distribution. 

\begin{itemize}
\item \textbf{Population Distribution}.
The distribution of nodes within the network is done according to the method seen in \cite{verbeek2003efficient, eckart2018hgmr, alam2015analysis}. It organizes population densities into a set of cluster locations and their respective Gaussians. Since these Gaussians are organized in a hierarchical model, there is a different solution for each level.
\item \textbf{Traffic Flow}.
Along with geographic location, our network utilizes an additional topology based on the distribution of messages within the space. Each node attempts to analyze outcomes from their own traffic, to update their own model. The solution space within our traffic-based routing model comprises of the set of clusters, and their respective Gaussians as well.
\end{itemize}

\subsection{Dual convex optimization.} 
Our network combines both of these network estimations to provide determinism as well as optimized for network load. The geographic density clustering ensures that clusters are connected to their neighbors, that messages will flow towards ever smaller radii so that corresponding hops are lower latency and that all nodes are connected within the network. While the geographic clustering is important to guarantee the integrity of the network, accommodations for message flow improve the effectiveness and efficiency of the network. 
 
Both estimation models can be incorporated through a dual convex optimization method, with the constraints forcing the solution space to contain a certain distribution of elements within clusters. Each node begins with a routing table distribution that mimics the geographic partitioning seen in \cite{pethalakshmi2014geo}. This routing table distribution is updated according to an objective function and a set of constraints. 
\begin{itemize}
    \item \textbf{Objective Function.} Each node seeks to construct its routing table to optimize its own individual objective function. This objective function incorporates a ranking factor for each peer within the table, and can include as features: path latency, availability, average latency, and load. 
        
    The objective function can additionally include features to lower the overall cost of communication. These features incorporate knowledge about message flow, as well as population density to reduce the number and distance of the hops that compose each message. 
    \item \textbf{Constraints.} The constraints within the function encapsulate aspects that are required for the network. These constraints are mainly updated to ensure that a node contains peers within the correct population distributions, and that nodes have enough peers within their tables. Constraints work to maintain the integrity of the network, as well as preventing overfitting.
\end{itemize}

\subsection{Validation and Convergence}
 This method of optimization is designed to converge towards the best solution for the specific network in which it is being employed. Gaussian distributions can be validated through a random sampling of probability distributions within the impacted area. According to the Law of Large Numbers, these distributions will  converge  to expected value even with a relatively small number of samplings. It is through this method that individual nodes can assess Gaussian Mixtures when they only have a sparse amount of information of the network. These updates also inform nodes about other solutions and enable convergence to the final solution.
 
 The convergence of both the Evolutionary Algorithm and the Particle Swarm Optimization Algorithm occurs relatively quickly compared to other optimization approaches. Further optimizations to the PSO methodology like those seen in \cite{zhan2009adaptive} can be used to improve convergence time in these models. 

\subsection{Overhead}
Our network does incur a continuous communication overhead in order to run its swarm optimizations. Both the clustering and the routing table optimization have the potential to utilize significant amounts of bandwidth. In order to reduce this cost, we can limit the number of updates. Communication overhead is represented as the expected amount of data that each nodes needs to send to another.

\begin{itemize}
    \item \textbf{Clustering.} The nodes send k data representing their distribution of clusters to log(N) nodes. The data broadcasts occur at a rate $t_{c}$, with $t_c$ eventually converging to 0.
    \item \textbf{Peer Table Optimization.}  The nodes send log(N) data representing a segment of their hash table to log(N) nodes. The data broadcasts occur at a rate $t_{p}$, with $t_p$ eventually converging to 0.
\end{itemize}

As stated previously, the eventual convergence of the problem will reduce the number of communications broadcast to 0. This means that while the network may incur an initially high communication cost, it will no longer incur this cost as the network grows older.

\section{Performance}

\subsection{Routing}
Routing within our network is influenced by the path length and the peer latency, which are used in combination to estimate the path latency within the network. Both path length and peer latency can be modeled in relation to the topology of the network, and the size of the network region.  It should be noted that these benchmarks represent the "worst case" of our network's routing capability because it is designed is to lower the weighted communication cost for each node, as it is related to the message distribution of individual nodes. 

\subsubsection{Path Length}
The path length represents the number of hops between any two nodes within the network, it is used to estimate the cost of communication, as well as the weighted path length within the network. Our network's path length is impacted by the number of levels within the cluster hierarchy's, the size of the clusters, and the number of clusters that are stored in each level. Despite this, all path lengths between distributed hash tables tend to share the same logarithmic bounds, with differences arising in the individual implementations themselves.

\begin{figure}[H]
\begin{tabular}{ |p{3.5cm}||p{3cm}|p{3cm}|p{3cm}|  }
 \hline
 \multicolumn{4}{|c|}{\textbf{Path Length}} \\
 \hline \hline
 &\textbf{Worst Case} & \textbf{Average Case} & \textbf{Best Case} \\
 \hline
 \textbf{Chord} & $log(N)$    & $log(N)$ &   1\\
 \textbf{ACO Chord} &   $log(N)$  & $log(N)$   & 1\\
  \textbf{Geo Chord} &   $log(N)$  & $log(N)$   & 1\\
 \textbf{Our Chord} & $log(N)$ &  $log(N)$ &  1 \\
 \textbf{Kademlia} & $log(N)$ &  $log(N)$&  1 \\
 \hline
\end{tabular}
\caption{In order to reduce complexity, it is assumed that the message being communicated has size 1, that the network has N nodes, and messages are being multicast to M nodes. }
\end{figure}

\subsubsection{Peer Latency}
 Along with path length peer latency is a major lever in reducing the weighted path length of messages within the network. While other networks use a random hashing scheme that does not account for distance within the network, ours uses geographic location and probability distributions to create latency bounds within different segments of the network. In fact, differences between peer latency within our network means that the latency of a hop within a cluster is lower than the latency of its parent cluster by a factor of k (with k being the number of sub-clusters). We model the average latency within each cluster using a rough estimation algorithm that assumes an equal distribution of nodes within each region, and integrates this Probability Density Function to calculate the average latency between two randomly selected points $(x_1, y_1)$ and $(x_2, y_2)$. \\
\begin{wrapfigure}{r}{0.4\textwidth} 
\centering
\begin{tikzpicture}
\begin{axis}[
    axis lines = left,
    xlabel = $Cluster$ $Levels$,
    ylabel = {$latency$ $(ms)$},
  height=5cm, width=5cm,
      title={\textbf{Peer Latency}},
    xmin=0, xmax=6,
    ymin=0, ymax=1200,
    xtick={0,1,2,3,4,5},
    ytick={0,200,400,600,800,1000},
    legend pos=outer north east,
    ymajorgrids=true,
    grid style=dashed,
]

\addplot [
    domain=0:6, 
    samples=20, 
    color=blue,
    ]
    {1000/(2^x)};
\addlegendentry{Cluster}
\addplot [
    domain=0:6, 
    samples=10, 
    color=red,
    dashed,
    ]
    {1000};
\addlegendentry{Average}
\addplot [
    domain=0:6, 
    samples=10, 
    color=black,
    ]
    {0};
\end{axis}
\end{tikzpicture}
\caption{Average latency within each neighborhood. Done with 2 clusters in each level.} \leavevmode \\
\begin{tikzpicture}
\begin{axis}[
    axis lines = left,
    xlabel = $Cluster$ $Levels$,
    ylabel = {$latency$ $(ms)$},
  height=5cm, width=5cm,
      title={\textbf{Path latency}},
    xmin=0, xmax=6,
    ymin=0, ymax=1200,
    xtick={0,1,2,3,4,5},
    ytick={0,200,400,600,800,1000},
    legend pos=outer north east,
    ymajorgrids=true,
    grid style=dashed,
]
\addplot [
    domain=0:6, 
    samples=20, 
    color=blue,
    ]
    {500/(2^x)};
\addlegendentry{Chord}
\addplot [
    domain=0:6, 
    samples=20, 
    color=blue,
    ]
    {600/(2^x)};
\addlegendentry{GeoChord}
\addplot [
    domain=0:6, 
    samples=20, 
    color=blue,
    ]
    {1000/(2^x)};
\addlegendentry{ACOChord}
\addplot [
    domain=0:6, 
    samples=10, 
    color=red,
    dashed,
    ]
    {1000};
\addlegendentry{Kademlia}
\addplot [
    domain=0:6, 
    samples=10, 
    color=black,
    ]
    {0};
\end{axis}
\end{tikzpicture}
\caption{Average latency within each neighborhood. Done with 2 clusters in each level.}
\end{wrapfigure}

\subsubsection{Path Latency}
As stated previously, the total latency of a path is directly related to the path length and the peer latency. When we combine the logarithmic path length with the exponential decrease in peer latency (as the clusters are sub-divided), we attain a network that has an anticipated path latency that is much smaller than that of traditional DHTs like Kademlia and Chord. The key to this difference in performance, is the decreasing hop latency that our network features, while traditional systems like Kademlia and Chord have the same expected latency with each hop.  
\subsection{Communication}
Our network utilizes the User Diagram Protocol (UDP) as a method of packet transmission with Forward Error Correcting to preserve data integrity. UDP offers scalable multi-casts, and can be used in browser clients through WebRTC. The cost of communication is equivalent to the number of nodes touched when data is transferred through the network. Path length is an important metric in peer-to-peer networks because it is equivalent to the cost of a search and a broadcast within the network. It should be noted that the UDP messages can be given responses, but flooded messages only demand responses if data is lost of corrupted. \\\\
A message can be sent to the entire network, a subset of the network, or a single node by flooding the message to their routing book. The communication overhead of this operation is equivalent to number of nodes touched (n) and the average path length in the network.

\begin{figure}[H]
\begin{tabular}{ |p{3.5cm}||p{2.5cm}|p{3.5cm}|p{3.5cm}|  }
 \hline
 \multicolumn{4}{|c|}{\textbf{Communication Overhead}} \\
 \hline \hline
 &\textbf{Worst Case} & \textbf{Average Case} & \textbf{Best Case (optimal)} \\
 \hline
 \textbf{Unicast} &  $log(N)$    & $log(N)$ &   1 \\
 \hline  
 \textbf{Multicast} &   M*$log(N)$  & M*$log(N)$   & $M$ \\
 \hline
 \textbf{Broadcast} &   N*$log(N)$  & N*$log(N)$   & $N$ \\
 \hline

\end{tabular}
\caption{In order to reduce complexity, it is assumed that the message being communicated has size 1, that the network has N nodes, and messages are being multicast to M nodes. }
\end{figure}

\subsection{Storage}
When compared to others like Kademlia and Geo-chord \cite{pethalakshmi2014geo}, our network has the potential to consume a greater amount of storage. While all of the distributed hash tables consume a logarithmic amount of storage, our network can add additional nodes to improve the routing effectiveness of the node itself.

Furthermore, much of the optimizations utilize additional variables to guide their decisions. The added storage of these additional constraints does not impact the theoretical size of the storage, but is an important aspect of the system to notice. 

\begin{figure}[H]
\begin{tabular}{ |p{4.5cm}||p{3cm}|  }
 \hline
  \multicolumn{2}{|c|}{\textbf{Storage Overhead}} \\
 \hline \hline
  \textbf{Our Network} & $log(N)$ + $M$  \\
 \textbf{Chord} & $log(N)$   \\
 \textbf{ACO Chord} &   $log(N)$  \\
  \textbf{Geo Chord} &   $log(N)$  \\
 \textbf{Kademlia} & $log(N)$  \\
 \hline
\end{tabular} 
\caption{In order to reduce complexity, it is assumed that the message being communicated has size 1, that the network has N nodes, and messages are being multicast to M nodes. }
\end{figure}

\subsection{Performance Optimizations}
One of the most beneficial aspects of utilising dual optimization for the agent update function is that requirements and features can be added in to maximize different network objective functions. Additionally, the network can be updated such that it can improve the overall effectiveness of the network. Performance optimizations include the caching of highly used paths, heuristic pruning of unused cluster connections, and a variety of heirarchical clustering methodologies. Each optimization comes at the cost of others, with the impacts shown in the table below.
\begin{figure}[H]
\begin{tabular}{ |p{3.5cm}||p{2.5cm}|p{2.5cm}|p{2.5cm}|
p{2.5cm}|}
 \hline
 \multicolumn{5}{|c|}{\textbf{Performance Relations}} \\
 \hline \hline
 &\textbf{Path Length} & \textbf{Peer Latency} & \textbf{Storage}  & \textbf{Path Latency}\\
 \hline
 \textbf{Path Length} & \cellcolor[RGB]{136, 204, 0}-    &  \cellcolor[RGB]{224, 224, 235}0 & \cellcolor[RGB]{255, 102, 102}+ &    \cellcolor[RGB]{136, 204, 0}- \\
 \hline  
 \textbf{Peer Latency}  & \cellcolor[RGB]{255, 102, 102}+    & \cellcolor[RGB]{136, 204, 0} - &  \cellcolor[RGB]{255, 102, 102}+ &   \cellcolor[RGB]{255, 102, 102}+ \\
 \hline
 \textbf{Storage}  & \cellcolor[RGB]{224, 224, 235}0   & \cellcolor[RGB]{255, 102, 102}+ &   \cellcolor[RGB]{136, 204, 0}- &   \cellcolor[RGB]{136, 204, 0}- \\
 \hline
\end{tabular}
\caption{In order to reduce complexity, it is assumed that the message being communicated has size 1, that the network has N nodes, and messages are being multicast to M nodes. }
\end{figure}

\begin{itemize}
    \item \textbf{Path Length}. The number of hops from node-to-node can be reduced by increasing the size of the routing paths such that the node has optimal paths to nodes that are closer.
    \item \textbf{Peer Latency}. All paths will feature a sub-optimal latency since they are not point to point. In order to reduce latency, the average number of hops must decrease. A reduction in peer latency does not necessarily decrease the latency of the path, but it does reduce the latency of each hop within the network. 
    \item \textbf{Storage}. Reducing storage costs, involves reducing the total network information that each peer receives. This translates to a lowered number of peers within the network. It should be noted that the size of the storage is inversely proportional to the path length, peer latency, and communication overhead. 
\end{itemize}

\section{Applications}
This scheme can be useful in networks that require messages to be flooded quickly, or require optimized messaging of similar data sets. Additional DHTs can be added to allow for better routing systems like Bittorrent \cite{johnsen2005peer} or IPFS \cite{benet2014ipfs}. 

Our proposed peer-to-peer scheme solves issues of efficiency, autonomy, and scalability that typical distributed systems face. It uses the routing methods of geochord \cite{pethalakshmi2014geo}, as well as the resilience of swarm optimization to deal with communication overhead and time as an underlying network. 

One of the advance of our network approach is that it acts as a very versatile underlying network that can be re-purposed to fit a variety of needs.

\begin{itemize}
\item \textbf{Network Synchronization}. 
Requires a flooding of the network, and can incur a high communication overhead for the network as a whole, but the overhead for each individual node is much lower. 
\item \textbf{File lookup/Messaging}. File systems must be included as a part of something extra. Additionally, it can route the file system based on geographic. Via pigeonholing Principle this can shown to have a shorter average latency than Kademlia. Can be similar to the chord routing system. However, it is likely that these will be separate for each use case. 
\item \textbf{Databases}. Easier domain routing and distribution for systems. Naive database queries are treated in much the same way a file searches, with the lookup of the data set and the execution of the query. The one optimization that databases, and in some cases file systems, is a cache that intelligently store and push data from the peers themselves.

\item \textbf{Pub-sub Data Streaming.}
While database querying is the same as file sharing, data streaming/pub-sub is a step up from database queries. Peers within the network can store a set of labels/or channels that they would like to participate in, either as a publisher or as a subscriber.

Connected to nodes. Location optimized, local minima from state positions w/ randomness to provide enhanced versatility. Streaming data can take place in parallel from multiple data sources in a seeding fashion similar to Bittorrent \cite{cohen2003incentives}. The data can be streamed from multiple devices, or multi-cast to many devices depending on the use case. In both cases, the data can be transmitted efficiently through UDP, with limited port overhead. 

\end{itemize}
\section{Conclusion}
As the prevalence of peer to peer networks grows, efficient networking schemes are becoming increasingly important. The current schemes of chord and Kademlia are unable to adequately account for proximity networks. Other schemes that utilize proximity-based routing do not create dynamic systems for the blockchain specific networks.

Our network uses proximity-based routing techniques seen in \cite{pethalakshmi2014geo}, along with dynamic updates based on multiple autonomous agents. Each peer within the network can act as its own agent. This creates a fault-tolerant, decentralized and efficient. Our scheme is set to be effective in cases where peers uses similar data streams, messages need to be propagated within the network, but can also be optimized and run underneath additional file name messaging protocols. These protocols can be accommodated by introducing additional requirements that meet the needs of the protocol. 
\begin{itemize}
    \item \textbf{Fault-tolerance}. Increasing the fault-tolerance of the system involves increasing the storage capacity of the address books to accommodate for a greater number of centroid nodes, to handle greater tolerance for nodes going offline.  
    \item \textbf{Precision}. In some cases, it helps to have greater control of clusters of nodes to accommodate for geographic load (e.g. geographic database storage allocation). In this case, it helps to increase the number of clusters, this will increase the number of peers each node holds which increases storage. Path length will also increase, but peer latency will lower. 
    \item \textbf{Relative Cluster Availability}. Average availability within each cluster. This is used to generate an expected number of available nodes within a cluster. Providing this metric enables clusters can be generated based on an estimated set of nodes, not situational network states.
\end{itemize}

Networking schemes are becoming increasingly useful in order to take the load off server-client communication. Further, these distributed networks put lower strain on each individual peer, provide agent-based autonomy, and provides an efficient method, to satisfy overlay networks requirements of autonomy, fault-tolerance, and scalability. 
Our proposed scheme can be tuned for a variety of objectives which presents a strong reason for it to be applied to file-sharing, distributed storage, and blockchain networks.

\bibliography{ms}

\end{document}